# Spatial separation of degenerate components of magnon Bose-Einstein condensate by using a local acceleration potential

I.V. Borisenko[1, 2,*], V.E. Demidov[1], V.L.Pokrovsky[3,4], and S. O. Demokritov[1]

[1]*Institute for Applied Physics and Center for Nanotechnology, University of Muenster, 48149 Muenster, Germany*

[2]*Kotel'nikov Institute of Radio Engineering and Electronics, Russian Academy of Sciences, 125009 Moscow, Russia*

[3] *Dept. of Physics and Astronomy, Texas A&M University, College Station, TX 77843-4242, USA*

[4] *Landau Institute of Theoretical Physics, Russian Academy of Sciences, Chernogolovka, Moscow Region 142432, Russian Federation.*

Bose-Einstein condensation (BEC) of magnons is one of few macroscopic quantum phenomena observable at room temperature [1-12]. Due to competition of the exchange and the magnetic dipole interactions the minimum-energy magnon state is doubly degenerate and corresponds to two antiparallel non-zero wavevectors [10]. Correspondingly, magnon BEC differs essentially from other condensates, since it takes place simultaneously at $\pm k_{min}$. The degeneracy of BEC and interaction between its two components have significant impact on the condensate properties. Phase locking of the two condensates causes formation of a standing wave of the condensate density and quantized vortices [11]. Additionally, interaction between the two components is believed to be important for stabilization of the **condensate with respect to the real-space collapse [12]. Thus, the possibility to create a non-degenerate, single-component condensate is decisive for understanding of underlying physics of magnon BEC. Here, we experimentally demonstrate an approach, which allows one to accomplish this challenging task. We show that this can be achieved by using a separation of the two components of the degenerate condensate in the real space by applying a local pulsed magnetic field, which causes their motion in the opposite directions. Thus, after a certain delay, the two clouds corresponding to different components become well separated in the real space. We find that the motion of the clouds can be described well based on the peculiarities of the magnon dispersion characteristics. Additionally, we show that, during the motion, the condensate cloud harvests non-condensed magnons, which results in a partial compensation of the condensate depletion**.

Bose-Einstein condensation (BEC) is one of the most striking manifestations of quantum nature of matter on the macroscopic scale. It represents a formation of a collective quantum state of particles with integer spin – bosons, at the minimum energy, if their density exceeds the critical one, strongly depending on temperature. Predicted by Einstein in 1925 theoretically [13], it was observed experimentally in diluted atomic alkali gases at ultra-low temperatures of $10^{-7}$ K [14,15]. Gases of quasi-particles are particularly attractive objects for the observation and study of BEC, because the conditions for BEC formation are not so extreme, as for real atoms. The effective masses of quasi-particles can be significantly smaller than those of atoms, leading to smaller critical densities and higher transition temperatures. Moreover, due to thermal fluctuations, the gases of quasi-particles possess large densities at non-zero temperatures, which can be further increased by using an external excitation. In the recent years, these advantages has led to the experimental observation of BEC of polaritons [16], excitons [17], and magnons [1].

BEC of magnons occupies a special place among quasi-particle BECs, since, up to now, this is the only BEC observed at room temperature. Since its discovery [1], significant efforts have been put into the studies of the nature of this phenomenon and understanding of characteristics of magnon condensates [2-9]. Very recently, such important issues as spatial and temporal coherence of the magnon condensate, its stability with respect to the collapse in the real space [12,18], the possibility of the superfluid behavior [8,19], as well as the existence of the second sound [20] and Bogolubov waves [21] have been addressed.

Another unique feature of the magnon BEC, which makes its study particularly interesting and challenging, is its double degeneracy. Since the magnon dispersion is governed by the competition of two interactions - the dipole interaction dominating at small wavevectors and the exchange interaction dominating at large wavevectors, - the spectrum of magnons in magnetic films possess two energy minima corresponding to non-zero wavevectors $\pm k_{\text{min}}$ [10]. Consequently, the magnon BEC is doubly degenerate: two condensate components are simultaneously formed in the two spectral minima at $k_{\text{BEC}}=\pm k_{\text{min}}$. Below we refer to these components described by the wave functions proportional to $\exp(+ik_{\text{min}}z)$ and $\exp(-ik_{\text{min}}z)$, as (+) and (−) components, respectively. We emphasize that, although the phase velocities of magnons at $\pm k_{\text{min}}$ can be as large, as several kilometers per second, the group velocity at these spectral points is zero. Therefore, the created condensate clouds do not move and occupy the same area in the real space.

The degeneracy has a significant impact on the properties of magnon BEC and on the interpretation of experimental results. For example, in [11], it was experimentally shown that the two components of the condensate can be phase-locked to each other, resulting in formation of a standing wave of the total condensate density in the real space. From the theoretical point of view, on one hand, it is believed that the interaction of the two components is important for the spatial stability of the entire condensate [12]. On the other hand, the same interaction is predicted to cause a symmetry breaking resulting in different densities of the two condensate components [7]. In order to address these phenomena in detail, it is necessary to elaborate an approach, which would allow one to create single-component magnon BEC experimentally. At first glance, the most straightforward approach could be based on utilization of magnon injection mechanisms breaking the symmetry between $+k$ and $-k$. However, all magnon pumping mechanisms used up to now are believed to satisfy intrinsically this symmetry, making the direct creation of single-component magnon BEC very challenging.

Here, we propose and experimentally test a different approach to the solution of this long-standing problem. Instead of the direct creation of single-component BEC, we implement a spatial separation of the two components of the condensate corresponding to $\pm k_{min}$. We first create a confined degenerate condensate cloud and then apply a pulse of localized magnetic field, which causes the two components of the condensate to move in the opposite directions. Thus, after a certain delay, the two clouds corresponding to different condensate components became well separated in the real space. We show that the observed velocities of the condensate clouds and their dependencies on the amplitude of the accelerating pulse are in a good agreement with theoretical values obtained from calculations of the magnon dispersion. Moreover, we find experimentally that the observed separation process conserves the total number of magnons in the split clouds.

The schematics of our experiment is illustrated in Fig. 1a. The magnon condensate is created at room temperature in yttrium iron garnet (YIG) film with the thickness of 5.1 μm and the lateral dimensions of 2×10 mm$^2$ by using microwave parametric pumping. The pumping field is produced by a dielectric resonator with the resonant frequency of $f_{MW} = 9$ GHz. The pumping injects primary magnons at the frequency $f_p = f_{MW}/2 = 4.5$ GHz, which thermalize and create BEC in the lowest-energy spectral states at $k_{BEC} = \pm k_{min}$. The frequency of BEC is mainly determined by the applied static magnetic field $H_0$ and is equal to 3.5 GHz at $H_0 = 1.2$ kOe. As discussed in detail below, in order to separate condensate clouds, they are put in motion by applying an additional localized field $\Delta H(z)$ produced by an electric current $I$ flowing through a control line with the width of 10 µm and the thickness of 400 nm, adjacent to the surface of YIG (Fig. 1a). Due to the shielding of the microwave field of the resonator by the metallic control line, the maximum pumping efficiency is reached in the area above the line. The microwave power applied to the resonator is chosen in such way that the BEC threshold is exceeded in the vicinity of the control line only. Thus, we create a BEC cloud with the width of about 50 µm along the $z$-axis. In the absence of the acceleration field, the cloud remains centered at the control line.

We study the condensate density and its spatial and temporal evolution using micro-focus Brillouin light scattering (BLS) technique [22,23] with the selectivity with respect to the wavevector of the scattering magnons [4,24,25]. As depicted in Fig. 1a, due to the conservation of the linear momentum, photons of the incident probing light are scattered by the (+) and (-) component of the condensate at different angles. Correspondingly, by selecting for the analysis one or the other part of the scattered beam (shown by different colors in Fig. 1a), one can address one or the other component.

Figures 1b and 1c illustrate the idea of the experiment. In order to put the condensate clouds in motion, we apply a pulse of additional localized magnetic field $\Delta H(z)$ with the rise time 5-10 ns. As a result, the initial condensate cloud is brought to the top of a potential hill defined by magnetic field profile $H_0 + \Delta H(z)$. Figure 1c shows the dispersion curves for magnons at $H_0$ and at $H_0 + \Delta H$. While the former curve corresponds to the magnon spectrum far from the control line, the latter illustrates the magnon spectrum close to its center. The frequency of the condensate above the control line is higher than that of the lowest-energy magnon state in the surrounding film. Experiencing the influence of the energy gradient caused by $\Delta H(z)$, the condensate cloud starts to move in the real space. As a result, according to the energy conservation law, the condensate evolves into states with the absolute value of the wavenumber, which is larger (exchange-dominated magnons) and/or smaller (dipole-dominated magnons) than $k_{BEC}$, as indicated in Fig. 1c by the dashed lines. Thus, two fractions (dipole- and exchange-dominated) are formed. Let us first consider the evolution of the (+) component of the condensate. It is obvious from Fig. 1c that, since the group velocity is determined by the slope of the dispersion curve, $v_{gr}=2\pi df/dk$, for a non-monotonic dispersion spectrum $f(k)$, two fractions of the (+) component move in the opposite directions. Moreover, since, even in the vicinity of $k_{BEC}$, $f(k)$ is strongly non-parabolic, the cloud of dipole-dominated magnons, corresponding to $k < k_{BEC}$ possesses a larger velocity than that built by exchange-dominated magnons, $k > k_{BEC}$. Thus, the (+) component is split in the real space into two clouds: a fast cloud moving in the (–z)-direction and a slow cloud moving into the (+z)-direction. By analyzing the motion of the (–) component, one obtains a fully symmetrical picture: its fast fraction moves into the (+z)-direction, whereas the slow one moves into (–z)-direction (see Fig. 1b).

Using the spatial and the temporal resolution of the BLS technique, we examine the above concept by recording the temporal and spatial profiles of the condensate density separately for the two components, as shown in Fig. 2. The data obtained for (+) and (–) component are shown as color-coded space-time maps in Figs. 2a and 2b, respectively. Note here that, since the horizontal axis in Figs. 2a and 2b shows the propagation distance, steeper trajectories in the maps correspond to a slower motion. Figures 2c and 2d illustrate spatial profiles of the density recorded at different time delays with respect to the start of the accelerating field pulse, as labelled. One can definitely see that, indeed, each component is split into two clouds moving in the opposite directions with different velocities. Moreover, the already mentioned symmetry connecting the direction of the wavevectors and that of the velocities is also clearly seen in Fig. 2. It is important to notice that the magnetic field induced by the current is strongly localized around the control line. Thus, the majority of the travel range shown in Fig. 2 is free from the field gradient, which allows us to consider the motion of condensate clouds as a free propagation.

To get deeper insight into the studied process, we plot in Fig. 3 the positions of the cloud centers as a function of time for both components. Red and blue colors correspond to (+) and (-) component, whereas circles and squares represent fast and slow fractions of the corresponding component. Dashed lines show linear fits of the experimental data. The slopes of the lines correspond to the velocity of the clouds. Figure 3 clearly demonstrates that, within the experimental accuracy, the clouds move with constant velocities: slow clouds move along the corresponding wavevectors $k_{\text{BEC}}$, whereas fast clouds move in the opposite direction, in agreement with the above discussion based on the magnon dispersion spectra shown in Fig. 1c. As also seen from Fig. 3, the trajectories of the clouds of (+) and (-) component start at $t=0$ at slightly different positions. The same can be seen, if one compares the one-dimensional traces in Figs. 2c and 2d for $t=0$: the center of the (+) component at $t=0$ is shifted in the positive direction, whereas the center of the (–) component is shifted in the negative one. We associate these small shifts with the motion of magnons in the real space during their thermalization and formation of the condensate. In fact, although the distribution of the pumping field is symmetric, non-zero group velocities of the primary and the intermediate magnons created as a result of the thermalization process can cause the observed shift.

Figure 4 illustrates the temporal evolution of the spatial width and of the number of magnons in the moving clouds. Figure 4a shows the variation of the width for the clouds formed by fast (dipole-dominated) and slow (exchange-dominated) magnons. $\Delta_0 = 50$ μm indicates the initial width of the cloud before application of the splitting field pulse. As seen from Fig. 4a, the splitting causes a constriction of both clouds: the width of the exchange-dominated cloud is smaller than $\Delta_0$ during the entire propagation interval, whereas that of the dipole-dominated cloud is smaller than $\Delta_0$ at $t < 100$ ns. Additionally, the moving clouds experience a dispersive broadening. For dipole-dominated magnons, this broadening is very strong, whereas for exchange-dominated magnons it is much weaker. These behaviors are consistent with the fact that the dispersion coefficient $D=2\pi d^2 f/dk^2$ of the dipole-dominated part of the magnon spectrum is much larger than that for the exchange-dominated part: $D_{\text{dip}}=3.9$ cm$^2$/(rad·s), $D_{\text{ex}}= 0.24$ cm$^2$/(rad·s) [15]. In agreement with this interpretation, the dashed straight lines representing the best linear fit of the experimental data with the ratio of two slopes equal to $D_{\text{dip}}/D_{\text{ex}}=16.2$, reproduce the observed broadening very well.

More intriguing is the evolution of the total number of magnons in the clouds, $N$, characterized in Fig. 4b. The shown data are the normalized integrals of the recorded spatial profiles of the condensate density versus propagation time for clouds formed by dipole-dominated, $N_{\text{dip}}$ and exchange-dominated magnons, $N_{\text{ex}}$, correspondingly. The data are normalized by the value of the integral for the initial condensate cloud existing before application of the splitting field pulse. Results of two types of experiments are presented in Fig. 4b. In the first case (open symbols), the microwave pumping was switched off at $t=0$. In the second case (solid symbols), the pumping was applied continuously. Keeping in mind the short thermalization time of magnons [2], the former case corresponds to the free propagation of the condensate clouds, whereas the latter case corresponds to the propagation of the condensate clouds interacting with thermalizing non-coherent magnons.

Before discussing the temporal decay of the magnon clouds, we would like to stress, that the data of Fig. 4 clearly indicate that the splitting process does not significantly reduce the total number of magnons in the condensate. Indeed, the sum $N_{dip} + N_{ex}$ remains close to $N_{tot}$ for $t<50$ ns. From the theoretical point of view, the main process changing the total number of magnons in the condensate cloud is the Cherenkov emission and absorption of condensate magnons by thermal magnons. However, this process is strongly suppressed under the conditions of the experiment due to a small Mach angle of the process, which results in a small statistical weight of final states.

In the case of the free propagation (open symbols in Fig. 4b), one observes a clear exponential decay of the number of magnons $N \sim \exp(-t/\tau)$ (note the log-linear scale of the graph) for both exchange-dominated and dipole-dominated magnons, although with slightly different decay times: $\tau_{ex}= 175$ ns, $\tau_{dip}= 230$ ns. In the case of the continuous pumping (solid symbols in Fig. 4b), the observed behaviors differ strongly. While the existence of non-condensed magnons practically does not influence the decay of the exchange-dominated condensate, $N_{dip}(t)$ changes dramatically. For $t<170$ ns (propagation distance $< 80$ µm) one observes a rather slow decay with the decay time $\tau^* = 290$ ns. This indicates an existence of a mechanism, which partially compensates the intrinsic decay of $N_{dip}$. We attribute these behaviors to the harvesting of non-condensed magnons by the cloud. As mentioned above, the control line modifies the pumping conditions around it, leading to a formation of a condensate cloud with the width of $\Delta_0=50$ µm (situated in the area ±25 µm around the center of the control line). Away from this area, the pumping strength is below the threshold of BEC. However, it is still strong enough to create an essential number of primary, and, as a consequence, of intermediate non-coherent magnons. A non-zero group velocity of the thermalizing magnons further increases the spatial extent of this zone. Thus, for some time, the accelerated condensate cloud propagates through the area, where these magnons are present. To understand the consequences, one should keep in mind, that the probability for a boson to scatter into a given spectral state is proportional to the population of this state. Outside of the condensation area, the density of non-coherent magnons is not high enough to create a condensate. However, as soon as the condensate cloud enters this area, the non-coherent magnons can efficiently scatter into the condensate and, in this way, partially compensate the decay of $N$. After some propagation time, however, the density of non-condensed magnons decreases, the harvesting effect becomes negligible, and the cloud again shows a fast decay. It is interesting to note, that, as follows from the experimental data, the exchange- and the dipole-dominated magnons are influenced differently by the harvesting process.

Finally, we discuss the effects of the amplitude of the accelerating magnetic field $\Delta H$, on the velocities of the clouds. Performing an analysis of the condensate movement at different $\Delta H$ values in the same way, as demonstrated in Fig. 3, we obtain the dependences of the velocities on $\Delta H$ (Fig. 5) for the dipole-dominated (circles) and exchange-dominated (squares) clouds. The solid curves in Fig. 5 present the results of calculations based on the magnon dispersion curves shown in Fig. 1c assuming that the velocities of the clouds are equal to the group velocities of magnons $v_{gr}=2\pi df/dk$ corresponding to the spectral states marked in Fig. 1c by circles. We emphasize that the results of calculations are in perfect agreement with the experimental data. This fact clearly indicates the validity of our interpretation of the observed behaviors.

In conclusion, we have experimentally demonstrated that doubly degenerate condensate of magnons can be split in the real space into two components ((+) and (–)), corresponding to the opposite wavevectors $\pm k_{\mathrm{BEC}}$, by applying a spatially localized pulse of magnetic field, which puts the components of the initially degenerate condensate in motion in the opposite directions. We find that this process is additionally accompanied by a splitting of each condensate component into two fractions constituted by dipole- and exchange-dominated magnons moving with different velocities. During the propagation, the two clouds evolve differently, as the broadening of the cloud and the reduction of the total numbers of magnons in the cloud are concerned. The faster, dipole-dominated condensate was found to harvest non-condensed magnons, which partially compensate the decay of the total number of magnons in the cloud. Our findings open a route for the efficient manipulation of magnon Bose-Einstein condensates in the real space and contribute to the deep understanding of the physics of degenerate condensates in general.

**Methods**

**Test system.** The magnon BEC was studied in a 5.1 µm-thick YIG film epitaxially grown on a Gadolinium Gallium Garnet (GGG) substrate. A strong microwave field necessary for magnon injection was created by using a dielectric microwave resonator with the resonant frequency $f_p$=9.055 GHz, fed by a microwave source. The device was placed into the uniform static magnetic field $H_0$=1.2 kOe. An additional spatially inhomogeneous magnetic field was created by using a Au control line carrying dc electric current (width 10 µm, thickness 400 nm, and length 8 mm). The line was fabricated on a sapphire substrate. The substrate was placed between the resonator and the YIG sample in such a way, that the conductor was directly attached to the surface of the YIG film (see Fig. 1a). The inhomogeneous field was applied in pulses with the duration of 1 μs and the period of 10 μs with the rise/fall times of 5-10 ns. Since the condensate clouds quickly leave the area of the localized field, the time of interaction between the field and the clouds was 10-50 ns, depending on the velocity of the cloud determined by the amplitude of the field pulse.

**Space, time- and k-resolved BLS measurements.** All the measurements were performed at room temperature. We focused the single-frequency probing laser light with the wavelength of 532 nm onto the surface of the YIG film through a transparent substrate and analysed the light inelastically scattered from magnons. The measured signal – the BLS intensity – is directly proportional to the magnon density. By varying the lateral position of the laser spot, we recorded spatial profiles of the magnon density. The integral of the BLS intensity over the cloud in the real space is proportional to the total number of magnons in the cloud. To analyze the spatio-temporal dynamics of the condensate, the BLS measurement was synchronized with the field pulse: the BLS signal was recorded as a function of the temporal delay with respect to the rising edge of the pulse. To achieve the wavevector selectivity, the scattered light beam was divided into two parts, as indicated by different colours in Fig. 1a. By obscuring one half of the scattered-light beam, we discriminated the BLS signal caused by one component of the condensate only.

**Data availability.** The data that support the findings of this study are available from the corresponding author upon reasonable request.

**References**

1. Demokritov, S. O. et al. Bose–Einstein condensation of quasi-equilibrium magnons at room temperature under pumping. *Nature* **443**, 430-433 (2006).

2. Demidov, V. E., Dzyapko, O., Demokritov, S. O., Melkov, G. A. & Slavin, A. N. Thermalization of a Parametrically Driven Magnon Gas Leading to Bose-Einstein Condensation. *Phys. Rev. Lett.* **99**, 037205 (2007).

3. Demidov, V. E., Dzyapko, O., Demokritov, S. O., Melkov, G. A. & Slavin, A. N. Observation of Spontaneous Coherence in Bose-Einstein Condensate of Magnons. *Phys. Rev. Lett.* **100**, 047205 (2008).

4. Demidov, V. E. et al. Magnon Kinetics and Bose-Einstein Condensation Studied in Phase Space. *Phys. Rev. Lett.* **101**, 257201 (2008).

5. Rezende, S. M. Theory of coherence in Bose-Einstein condensation phenomena in a microwave-driven interacting magnon gas. *Phys. Rev. B* **79**, 174411 (2009).

6. Serga, A. A. et al. Bose–Einstein condensation in an ultra-hot gas of pumped magnons. *Nat. Commun.* **5**, 3452 (2013).

7. Li, F., Saslow, W. M. & Pokrovsky, V. L. Phase Diagram for Magnon Condensate in Yttrium Iron Garnet Film. *Sci. Rep.* **3**, 1372 (2013).

8. Sun, C., Nattermann, T. & Pokrovsky, V. L. Unconventional Superfluidity in Yttrium Iron Garnet Films. *Phys. Rev. Lett.* **116**, 257205 (2016).

9. Dzyapko, O. et al. High-Resolution Magneto-Optical Kerr-Effect Spectroscopy of Magnon Bose–Einstein Condensate. *IEEE Magn. Lett.* **7**, 3501805 (2016).

10. Kalinikos, B. A. & Slavin, A. N. Theory of dipole-exchange spin-wave spectrum for ferromagnetic films with mixed exchange boundary conditions. *J. Phys. C* **19**, 7013–7033 (1986).

11. Nowik-Boltyk. P., Dzyapko, O., Demidov, V. E., Berloff, N. G. & Demokritov, S. O. Spatially non-uniform ground state and quantized vortices in a two-component Bose-Einstein condensate of magnons. *Sci. Rep.* **2**, 482 (2012).

12. Dzyapko, O. et al. Magnon-magnon interactions in a room-temperature magnonic Bose-Einstein condensate. *Phys. Rev. B* **96,** 064438 (2017).

13. Einstein, A. Quantentheorie des einatomigen idealen Gases, 2. Abhandlung. *Sitzungsberg. Ber. Preuss. Akad. Wiss.* **1,** 3 (1925).

14. Anderson, M. H. et al. Observation of Bose-Einstein condensation in a dilute atomic vapor. *Science* **269,** 198 (1995).

15. Davis, K. B. et al. Bose-Einstein condensation in a gas of sodium atoms. *Phys. Rev. Lett.* **75,** 3969 (1995).

16. Kasprzak, J. et al. Bose–Einstein condensation of exciton polaritons. *Nature* **443**, 409–414 (2006).

17. Eisenstein, J. & MacDonald, A. Bose–Einstein condensation of excitons in bilayer electron systems. *Nature* **432** 691–694 (2004).

18. Borisenko, I. V. et al. Direct evidence of spatial stability of Bose-Einstein condensate of magnons. Preprint at https://arxiv.org/abs/1910.06013 (2019).

19. Takei, S. & Tserkovnyak, Y. Superfluid spin transport through easy-plane ferromagnetic insulators. *Phys. Rev. Lett.* **112**, 227201 (2014).

20. Tiberkevich, V. et al. Magnonic second sound. *Sci. Rep.* **9**, 1 (2019).

21. Bozhko, D. A. et al. Bogoliubov waves and distant transport of magnon condensate at room temperature. *Nat. Commun.* **10**, 2460 (2019).

22. Demokritov, S.O. & Demidov, V.E. Micro-Brillouin Light Scattering Spectroscopy of Magnetic Nanostructures. *IEEE Trans. Mag.* **44**, 6 (2008).

23. Demidov, V. E. & Demokritov, S. O. Magnonic Waveguides Studied by Microfocus Brillouin Light Scattering. *IEEE Trans. Mag.* **51**, 0800215 (2015).

24. Neumann, T., Schneider, T., Serga, A.A. & Hillebrands, B. An electro-optic modulator-assisted wavevector-resolving Brillouin light scattering setup. *Rev. Sci. Instrum.* **80,** 053905 (2009).

25. Madami, M. et al. Direct observation of a propagating spin wave induced by spin-transfer torque. *Nat. Nanotech.* **6**, 635–638 (2011).

**Corresponding author:** Correspondence and requests for materials should be addressed to I.V.B. (boriseni@uni-muenster.de).

**Acknowledgements**

**Author Contributions:** I.V.B. and V.E.D. performed measurements, I.V.B, S.O.D,.and V.L.P performed data analysis, S.O.D. formulated the idea of the experiment and managed the project. All authors co-wrote the manuscript.

**Competing Interests:** The authors declare no competing interests.

**Figures and figure legends**

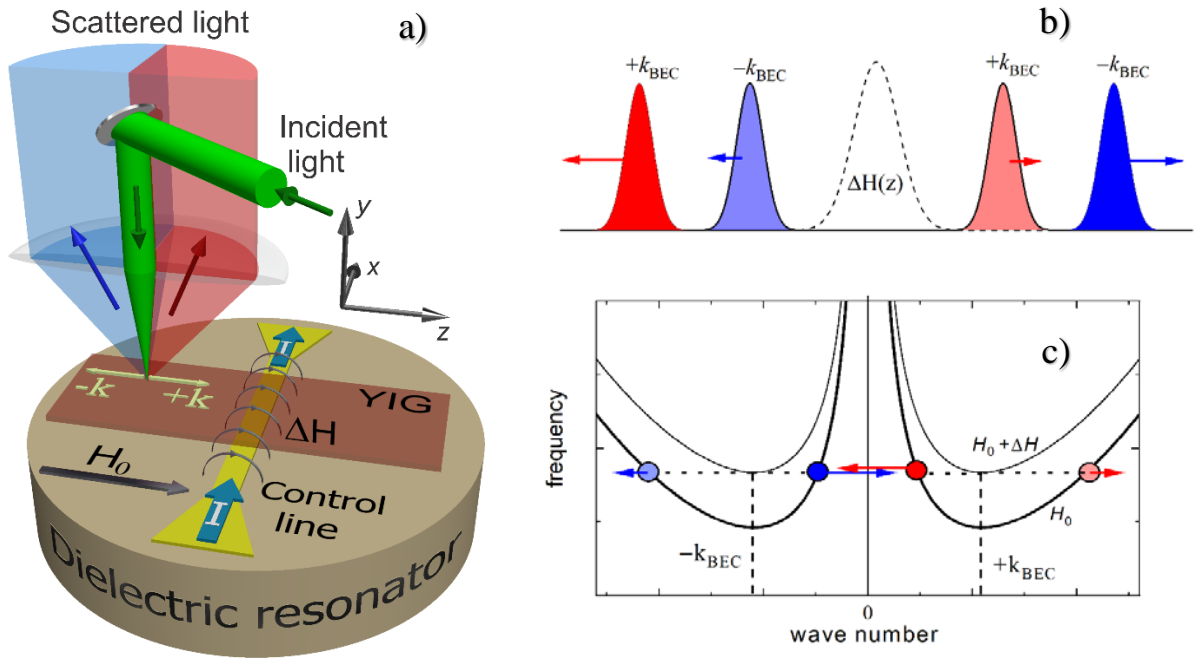

**Fig. 1 | Idea of the experiment. a,** Schematic view of the experimental system. Dielectric resonator creates a microwave-frequency magnetic field, which parametrically excites primary magnons in the YIG film. dc electric current, $I$ in the control line placed between the resonator and the YIG film, produces a localized magnetic field, $\Delta H$ which adds to the uniform static field $H_0$. The local density of condensed magnons is recorded by BLS with the probing laser light focused onto the surface of the YIG film. Scattering direction of light depends on the direction of the magnon wavevector enabling spatial separation of the corresponding BLS signals. **b,** After a pulse of localized magnetic field is applied, BEC cloud splits into four sub-clouds each of them moving in the direction, which is determined by the direction of the wavevector of the condensate and by the corresponding magnon spectral branch. **c,** Magnon dispersion curves corresponding to the center of the metal strip, where magnetic field is equal to $H_0+\Delta H$, and to a distant point, where the magnetic field is equal to $H_0$. Colored circles mark spectral states corresponding to the four split BEC clouds after they move away from center point. Arrows schematically indicate direction of the corresponding group velocities in the real space.

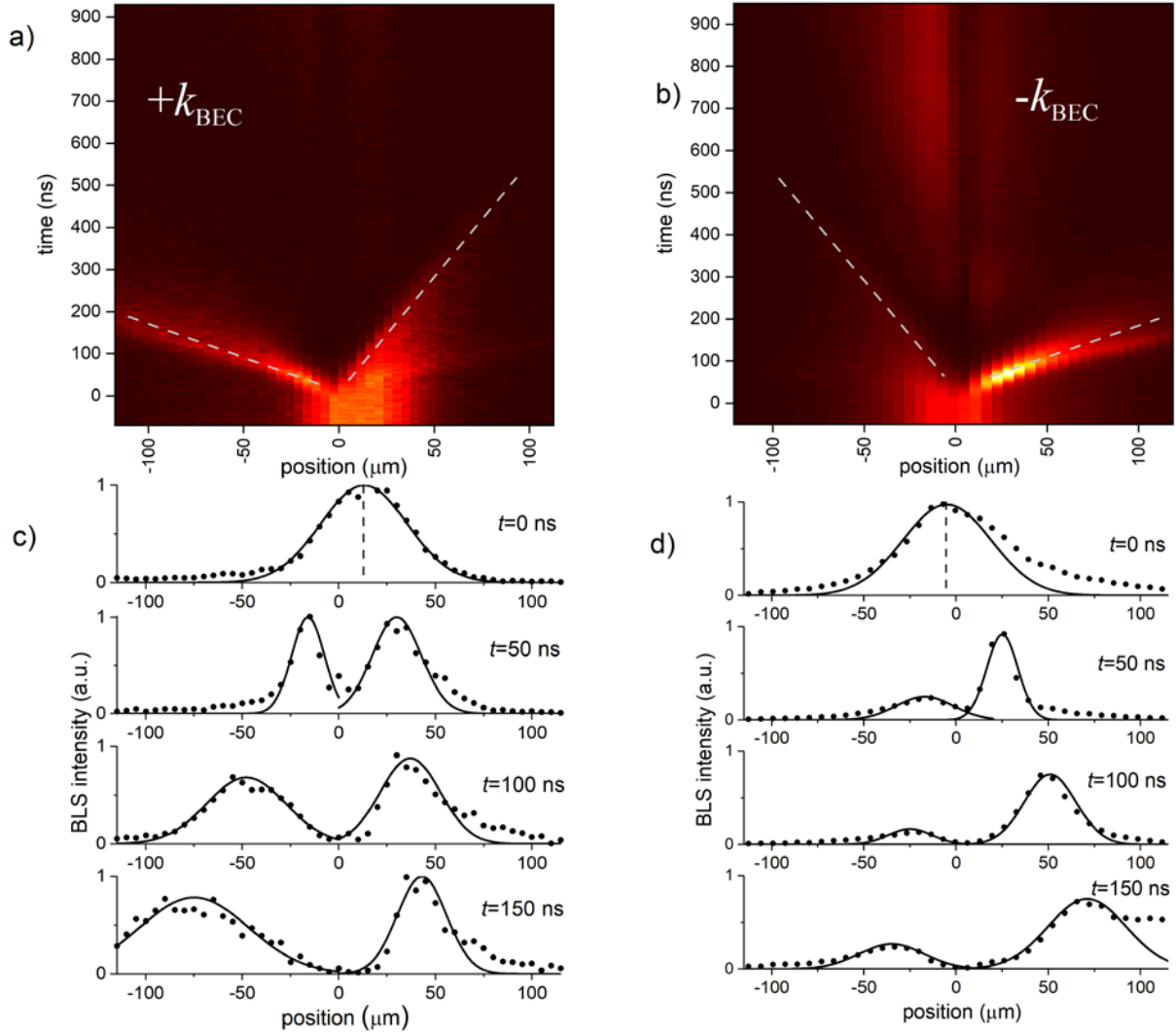

**Fig. 2 | Evolution of BEC under influence of magnetic field pulse. a, b,** Color-coded maps presenting the intensity of the BLS signal in time-space coordinates recorded at $H_0 = 1.2$ kOe and $\Delta H = 25$ Oe for (+) and (-) components of the condensate, respectively. The maps illustrate the motion of the corresponding clouds after the accelerating field is applied at $t=0$. "Fast" and "slow" rays propagating in the opposite directions are clearly seen in the maps. Dashed lines are guides for the eye. **c, d,** Representative normalized spatial profiles of the clouds recorded at different delay times, as labelled. On can see how the initially broad peak at $t=0$ splits into two narrow subpeaks travelling with different velocities. Dots are experimental data. Curves show the best fits by a peak function.

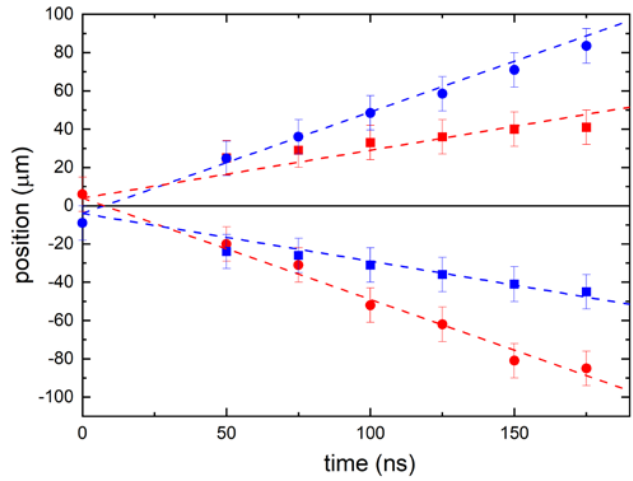

**Fig. 3 | Splitting of BEC clouds.** Temporal dependences of positions of the four BEC clouds formed after a magnetic-field pulse $\Delta H = 25$ Oe is applied at $t=0$. Red and blue colors represent (+) and (−) components of the condensate, respectively. Circles and squares correspond to "fast" and "slow" clouds, respectively. Dashed lines are best linear fits to the data. $H_0 = 1.2$ kOe. Error bars are SEM.

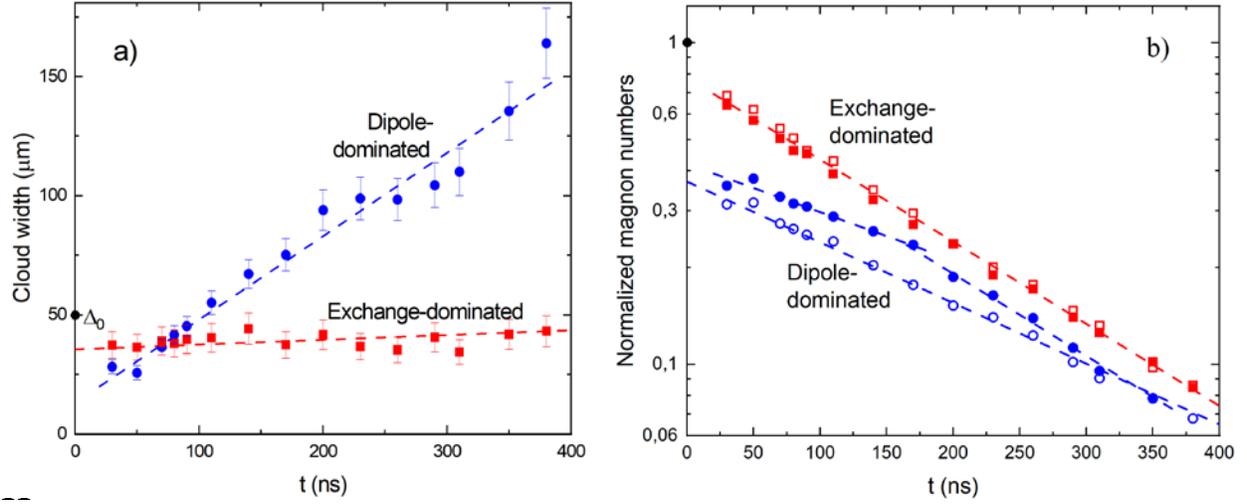

**Fig. 4 | Evolution of BEC clouds during propagation. a,** Variation of the spatial widths of the clouds versus propagation time, $t$, for the dipole-dominated (circles) and the exchange-dominated (squares) clouds for $\Delta H = 25$ Oe and $H_0 = 1.2$ kOe. $\Delta_0 = 50$ μm indicates the initial width of the condensate cloud before the splitting takes place. Dashed lines represent the best linear fit of the experimental data with the ratio of the two slopes equal to the ratio of the corresponding dispersion coefficients $D_{dip}/D_{ex}=16.2$. Error bars are SEM. **b,** Normalized magnon numbers in the condensate clouds versus $t$. Open symbols show the data obtained without microwave pumping, whereas solid symbols describe the experiment, where the pumping was continuously applied resulting in a continuous injection of non-condensed magnons. Dashed lines are exponential fits of the experimental data. Note a crossover from the slow to the fast decay for the dipole-dominated cloud in the presence of the pumping.

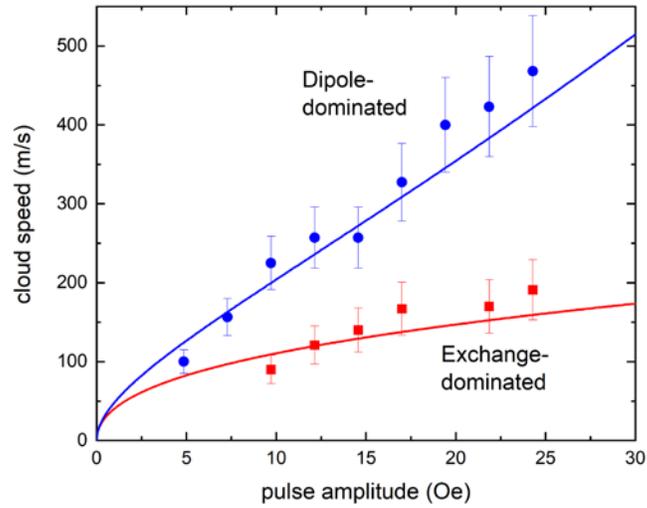

**Fig. 5 | Velocity of the condensate clouds.** Dependence of the cloud velocity on the magnitude of the accelerating magnetic field pulse $\Delta H$ at $H_0$ = 1.2 kOe. Blue circles and red squares correspond to the "fast" dipole-dominated and "slow" exchange-dominated clouds, respectively. Solid curves show the results of calculations of the magnon group velocity for the spectral states corresponding to the accelerated clouds (see Fig. 1c): the blue curve corresponds to the dipole part of the spectrum, whereas the red curve corresponds to the exchange part of the spectrum. Error bars are SEM.